\documentclass[aps,pra,superscriptaddress,showpacs,twocolumn]{revtex4-1}

\usepackage{amsmath,amssymb,graphicx,color,bm,soul,ulem}
\usepackage[english]{babel}
\usepackage[dvipsnames]{xcolor}
\usepackage{soul}
\usepackage[colorlinks=true,citecolor=blue]{hyperref}
\hypersetup{colorlinks=true,citecolor=blue,linkcolor=blue,urlcolor=blue}

\begin{document}

\title{Geometric Interpretation of Sum Photon Blockade}

\author{T. A. Khudaiberganov}
\affiliation{Department of Physics and Applied Mathematics, Vladimir State University named after A. G. and N. G. Stoletovs, 87 Gorkii st., 600000 Vladimir, Russia}

\date{\today}

\begin{abstract}
We present a geometric interpretation of the sum photon blockade effect in multimode quantum optical systems, such as semiconductor microresonators. The blockade condition \(c^{(n)} \cdot v = 0\) reflects the orthogonality of the \(n\)-photon amplitude vector to a target mode vector in an \(N\)-dimensional Hilbert space, visualized as the confinement of the state to a hyperplane.

A key result is the calculation of the maximum probability of the system remaining in the blockade subspace under the influence of decoherence processes (in particular, dephasing), which determines the practical feasibility and robustness of the effect. This approach extends to higher-order correlators \(g^{(2)}_\Sigma\) and cross-correlations, enabling the design of scalable quantum devices.

We introduce the concept of "dark-state typicality": as the number of modes \(M\) increases, the dark subspace annihilated by the collective mode operator asymptotically occupies a unit fraction of the \(n\)-boson Hilbert space. This allows the transition from fragile, finely tuned mechanisms to macroscopically robust non-classical light in large multimode bosonic architectures.

We consider continuum collective modes, hypotheses on correlation zeros and invariant manifolds, as well as the relationship between blockade and entanglement. 


\end{abstract}

\maketitle

\section{Introduction}

In modern quantum optics, the generation of non-classical light with sub-Poissonian photon statistics faces a fundamental scalability barrier [1]. Existing approaches that utilize photon blockade mechanisms [2, 3] require rigid, fine-tuned parameters. Generally, the blockade of two-photon emission relies on the fact that a second pump photon cannot find an energetically allowed state within the discrete spectrum. This is achieved via energetic anharmonicity, where the detuning of the second energy level relative to the first exceeds the first level's excitation from the vacuum by an amount proportional to the nonlinearity, which must be significantly larger than the line width [2]. In contrast, the unconventional mechanism relies on precise destructive quantum interference between different excitation pathways [3, 4]. As the number of emitters or modes grows, maintaining the conditions necessary for sub-Poissonian statistics becomes increasingly difficult: any minor parameter mismatch or external pure dephasing rapidly destroys the delicate interference balance required for quantum antibunching [5].

This problem is deeply rooted in the underlying structure of the Hilbert space. According to the principle of quantum typicality, the overwhelming majority of states in high-dimensional systems—referred to as "typical" states—exhibit statistical properties close to classical or thermal ensembles. Methods like UPB essentially attempt to trap the system in a rare, "anti-typical" configuration. However, as the number of modes increases, the "curse of dimensionality" takes over: local noise and dephasing inevitably drive the system out of this fragile manifold into the dominant volume of "bright" and noisy states where the blockade vanishes. Within this paradigm, sub-Poissonian statistics remain a rare exception requiring constant active control.

In this work, we propose a paradigm shift to resolve this tension. We introduce the concept of \textit{dark-state typicality}: as the number of modes $M$ grows, the dark subspace annihilated by the collective mode operator $A^n$ asymptotically occupies a unit fraction of the $n$-boson Hilbert space $\mathcal{H}_n$. 

This geometric concentration establishes a clear physical watershed between state-engineering strategies. While individual-emitter correlations in UPB succumb to the volume of the bright sector, collective blockade naturally exploits the topology of the high-dimensional state space. The collectively dark sector acts not as an isolated sanctuary for fine-tuned configurations, but as a generic attractor for the typical dynamics of the system [6]. Consequently, shifting the focus to collective sum modes allows us to transform the generation of non-classical light from a fragile, fine-tuned effect into an inherently robust, macroscopic property of large multimode bosonic architectures.

\section{Theoretical framework}


In a multimode bosonic system with $M$ input modes, the sum mode operator is defined as
\begin{equation}\label{eq:sum_mode}
\hat{A} = \boldsymbol{\alpha} \hat{\mathbf{a}},
\end{equation}
where $\hat{\mathbf{a}} = (\hat{a}_1, \dots, \hat{a}_M)^\top$ is the vector of annihilation operators for the input modes ($[\hat{a}_i, \hat{a}_j^\dagger] = \delta_{ij}$), and $\boldsymbol{\alpha} \in \mathbb{C}^M$ is the vector of complex superposition coefficients (usually normalized as $\|\boldsymbol{\alpha}\|^2 = 1$). Such a sum mode can be obtained using a passive linear optical (PLO) system described by a unitary matrix $U$, where $\hat{A}$ is realized as an output mode: $\boldsymbol{\alpha}^\dagger = \mathbf{u}^\dagger$ is a row of $U$. 

The second-order correlation function for this sum mode is given by
\begin{equation}\label{eq:second_order}
g^{(2)}_\Sigma = \frac{\langle \hat{A}^{\dagger 2} \hat{A}^2 \rangle}{\langle \hat{A}^\dagger \hat{A} \rangle^2},
\end{equation}
where the expectation values \(\langle \cdot \rangle\) are taken with respect to the system's state \(|\psi\rangle\). The numerator \(\langle \hat{A}^{\dagger 2} \hat{A}^2 \rangle\) quantifies the probability of detecting photon (or boson) pairs and can be expanded by considering the action of \(\hat{A}^2\) on the two-particle component of the state.

Assume the state is truncated to at most two particles per mode for simplicity (higher occupations would require additional terms, but this is sufficient for low-density regimes common in quantum optics):
\begin{equation}\label{eq:wave_function}
|\psi\rangle = |\psi^{(0)}\rangle + |\psi^{(1)}\rangle + |\psi^{(2)}\rangle + \cdots,
\end{equation}
where \(|\psi^{(n)}\rangle\) denotes the \(n\)-particle subspace. The operator \(\hat{A}^2\) annihilates two particles, so it acts non-trivially only on \(|\psi^{(2)}\rangle\), mapping it to the vacuum \(|0\rangle\) with a complex amplitude. Specifically,
\begin{equation}\label{eq:two_particle_wave_function}
|\psi^{(2)}\rangle = \sum_{i=1}^M c_{ii} |2_i\rangle + \sum_{1 \leq i < j \leq M} c_{ij} |1_i, 1_j\rangle,
\end{equation}
where \(c_{ii}\) is the complex amplitude for the state with two indistinguishable bosons in mode \(i\) (i.e., \(|2_i\rangle = \frac{\hat{a}_i^{\dagger 2}}{\sqrt{2}} |0\rangle\)), and \(c_{ij}\) is the amplitude for one boson in mode \(i\) and one in mode \(j\) (i.e., \(|1_i, 1_j\rangle = \hat{a}_i^\dagger \hat{a}_j^\dagger |0\rangle\)). 

The numerator \eqref{eq:second_order} then is:
\begin{equation}
\langle \hat{A}^{\dagger 2} \hat{A}^2 \rangle \approx \left| \sum_{i=1}^M \sqrt{2} \, \alpha_i^2 c_{ii} + \sum_{1 \leq i < j \leq M} 2 \alpha_i \alpha_j c_{ij} \right|^2.
\label{eq:correlator_of_sum_blockade}
\end{equation}




Consider the two-photon amplitude vector \(\mathbf{c}^{(2)}\), whose components are the complex coefficients \(c_{ij}\) from the decomposition of the two-particle state see \eqref{eq:two_particle_wave_function}
The vector \(\mathbf{c}^{(2)}\) nested in an \(N\)-dimensional complex space, where \(N = \binom{M+1}{2}\) is the dimension of the symmetric two-particle subspace for \(M\) modes. 
The target mode vector \(\mathbf{v}\) encodes the desired superposition weights in the sum mode, with components \(v_{ii} = \alpha_i^2\), and for distinct modes, \(v_{ij} = 2 \alpha_i \alpha_j\).

The sum blockade condition is given by the vector dot product

\begin{equation}
\mathbf{c}^{(2)} \cdot \mathbf{v} = 0,
\label{eq:sum_blockade_condition}
\end{equation}

here the dot denotes the scalar product of vectors and \(\mathbf{v}\) is the complex vector

\[\mathbf{v} =\sqrt{2} \left( \alpha_1^2, \alpha_2^2, \dots,  \alpha_M^2, \, \sqrt{2}\alpha_1 \alpha_2, \sqrt{2}\alpha_1 \alpha_3, \dots, \sqrt{2}\alpha_{M-1} \alpha_M \right),\]

where the first \(M\) components correspond to the double-occupancy terms \(v_{ii}\) and the remaining \(\binom{M}{2}\) to the inter-mode pair terms \(v_{ij}\) for \(i < j\). 

It is clear that there will always exist states satisfying the total blockade condition \eqref{eq:sum_blockade_condition}, i.e., I can make any quantum state blocked by specially choosing the parameters $\alpha_i$. Moreover, unlike other types of quantum blockade (such as unconventional blockade), there will be a multitude of such states. As the number of modes \(M\) increases, the fraction of the subspace dimension satisfying the total blockade condition, \(\frac{d-1}{d} = 1 - \frac{2}{M(M+1)}\), approaches 1. This means that in large spaces (for large \(M\)), the "proportion" of such states becomes closer to the full measure of the space.

If the system exhibits total quantum blockade on G1, G2, and on \( G = G_1 \cup G_2 \), then the coherent sum of the two-particle state amplitudes distributed between G1 and G2 equals zero:


\begin{equation}\label{eq:inter-group}
\sum_{k \in G_1} \sum_{l \in G_2} \alpha_k \alpha_l c_{\vec{\delta}_k + \vec{\delta}_l} = 0,
\end{equation}

where $c_{\vec{\delta}_k + \vec{\delta}_l}$ are the complex amplitudes of two-particle states with one boson in mode $k \in G_1$ and one in mode $l \in G_2$ (here, $\vec{\delta}_k$ denotes the standard basis vector for a single-particle excitation in mode $k$, such that $|\vec{\delta}_k + \vec{\delta}_l\rangle = \hat{a}_k^\dagger \hat{a}_l^\dagger |0\rangle$. 

The joint constraints for blockades on graphs \(G_1\), \(G_2\), and \(G_1 \cup G_2\) impose additional restrictions on quantum states. However, the fraction of the subspace satisfying criterion \eqref{eq:inter-group} also increases with the number of modes. If we have groups $G_1$, $G_2$ and the inter-group blockade condition \eqref{eq:inter-group}, then for a fixed quantum state there exists a choice of parameters ${\alpha_k \mid k \in G_1 \cup G_2}$ satisfying all three blockade conditions (on $G_1$, $G_2$, and $G_1 \cup G_2$) provided that the total number of modes $|G| = |G_1 \cup G_2|$ is sufficiently large.

\subsection{Engineering Entanglement via the Sum Blockade Effect}


A more convenient criterion for entanglement in certain multimode scenarios is the entanglement witness—a Hermitian operator W for which (under entanglement condition) $\operatorname{Tr}(W \rho) < 0$, since this approach directly probes specific correlations without requiring full state tomography of $\rho$. 

Hillery and Zubairy (HZ) derived an entanglement witness for continuous-variable systems, particularly for angular momentum operators in bosonic modes (e.g., photons or atoms in optical lattices). For a bipartition into disjoint subgraphs \(G_1\) and \(G_2\) (with \(G_1 \cap G_2 = \varnothing\), ensuring no overlapping modes), the HZ criterion is expressed in terms of collective number and transfer operators:
\begin{equation}\label{eq:collective_operators}
\hat{A}_{G_1} = \sum_{k \in G_1} \alpha_k\hat{a}_k, \quad \hat{A}_{G_2} = \sum_{l \in G_2} \alpha_l\hat{a}_l,
\end{equation}
The criterion takes the follow form:
\begin{equation}\label{eq:CEHZ_criterion}
\operatorname{CEHZ}^{(G_1, G_2)} = \frac{\langle \hat{A}_{G_1}^{\dag}\hat{A}_{G_1} \hat{A}_{G_2}^{\dag} \hat{A}_{G_2} \rangle}{|\langle \hat{A}_{G_2}^\dagger \hat{A}_{G_1} \rangle|^2} \stackrel{\rm ent}{\ll} 1,
\end{equation}
where \(\langle \cdot \rangle = \operatorname{Tr}(\rho \cdot)\) denotes the expectation value in the state \(\rho\). For separable states, \(\operatorname{CEHZ} \geq 1\) 
, while \(\operatorname{CEHZ} < 1\) certifies entanglement. 




Under the condition that \( G_1 \cap G_2 = \varnothing \), i.e., there are no common elements when partitioned into two subgraphs $G_1$ and $G_2$, the numerator in the Hillery-Zubairy (CEHZ) criterion Eq.~\eqref{eq:CEHZ_criterion} can be written as:

\begin{equation}\label{eq:two-particle_amplitudes}
\langle \hat{A}_{G_1}^{\dag}\hat{A}_{G_1} \hat{A}_{G_2}^{\dag} \hat{A}_{G_2} \rangle \approx \left| \sum_{k \in G_1} \sum_{l \in G_2} \alpha_k \alpha_l c_{\vec{\delta}_k + \vec{\delta}_l} \right|^2.
\end{equation}

Assuming the denominator in Eq.~\eqref{eq:CEHZ_criterion} remains finite and positive ($|\langle \hat{A}_{G_2}^\dagger \hat{A}_{G_1} \rangle|^2 > 0$, indicating non-zero single-particle coherences between the groups), the HZ criterion yields $\operatorname{CEHZ}^{(G_1, G_2)} \approx 0$. It then follows from Eqs.~\eqref{eq:inter-group} and~\eqref{eq:two-particle_amplitudes} that the complete blockade on graphs $G_1$, $G_2$, and $G$ induces entanglement between the angular momenta of the sum modes of graphs $G_1$ and $G_2$.  
Note that the reverse proposition generally does not hold true. The presence of entanglement in a system does not guaranty blockade behavior.

The discovered relationship between blockade and entanglement provides a practical method for the creation of engineered multimode entanglement in quantum networks. 

\begin{figure}\label{fig1}
\center{\includegraphics[width=\columnwidth]{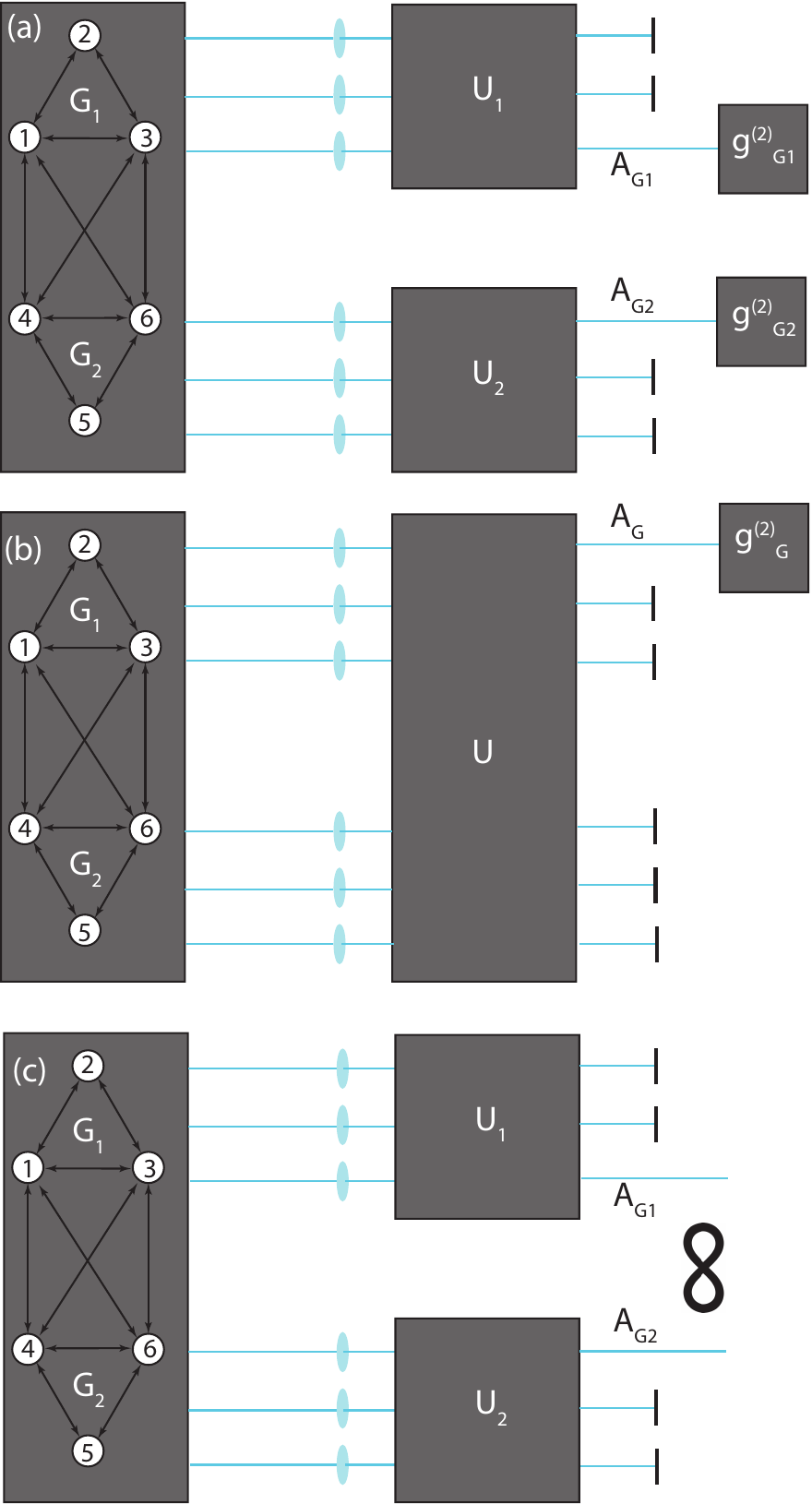}}
\caption{Scheme. 
}
\end{figure}

\section{Continuum Collective Mode}

We can define for  continuous case as a \emph{continuum collective mode}. Unlike a discrete collective excitation, which is constructed from a finite number of discrete modes, a continuum collective super-mode is defined directly within a continuous Hilbert space and is regarded as a primary geometric object.

Let $\Omega$ be a continuous domain of parameters and let $a(x)$ denote the bosonic
annihilation operator associated with the point $x\in\Omega$. A
continuum collective mode is defined by

\begin{equation}
\hat{A}[\phi]
=
\int_\Omega
\phi(x)\hat{a}(x)\,dx,
\end{equation}

where the mode function $\phi\in L^2(\Omega)$

is normalized according to

\begin{equation}
\int_\Omega
|\phi(x)|^2dx
=
1.
\end{equation}

The function $\phi(x)$ determines how elementary microscopic modes are
coherently distributed over the entire Hilbert space. Different choices
of $\phi(x)$ correspond to  unitary
mode rotation.

\subsection{Correlation zeros and correlation blockade}

For the continuum operators (11), we can define families of correlators
and the corresponding conditions for their zeros:

Let $B[\phi]
=
\int_\Omega
\phi(x)a(x)\,dx$
be a continuum collective mode, and
$
\mathcal K
=
\{K_\alpha(B,B^\dagger)\}
$
be a chosen family of operator-valued correlation functionals,
written in normal order.

A density operator $\rho$ is said to possess a \emph{correlation zero}
with respect to the operator $K_\alpha$ if

\begin{equation}
\mathrm{Tr}
\left(
\rho K_\alpha
\right)
=
0.
\end{equation}

The corresponding operator $K_\alpha$ is called a
\emph{vanishing correlation invariant} of the given state.

The set of all correlation zeros fully characterizes
the correlation structure of the state with respect to the chosen family
of observables.







\textbf{Hypothesis} (Continuum Correlation Blockade). For a continuum collective mode, the set of states satisfying a prescribed Correlation Blockade forms an invariant (or asymptotically invariant) manifold of the Liouvillian dynamics. Moreover, in the high-dimensional limit, concentration of measure suppresses fluctuations of the correlation invariants, making the Blockade robust against sufficiently weak collective and local perturbations.

\textbf{Hypothesis} (Discretization Leakage). Let $M_{inf}$ denote the Correlation Manifold of a continuum collective mode. Its finite-dimensional approximation $M_N$ obtained by projection onto N modes generally fails to preserve all correlation constraints. Consequently, parts of the continuum manifold become inaccessible, producing finite-dimensional gaps ("holes") in the Correlation Landscape. The size of these holes decreases as N increases and vanishes in the continuum limit.

\begin{figure}\label{fig1b}
\center{\includegraphics[width=\columnwidth]{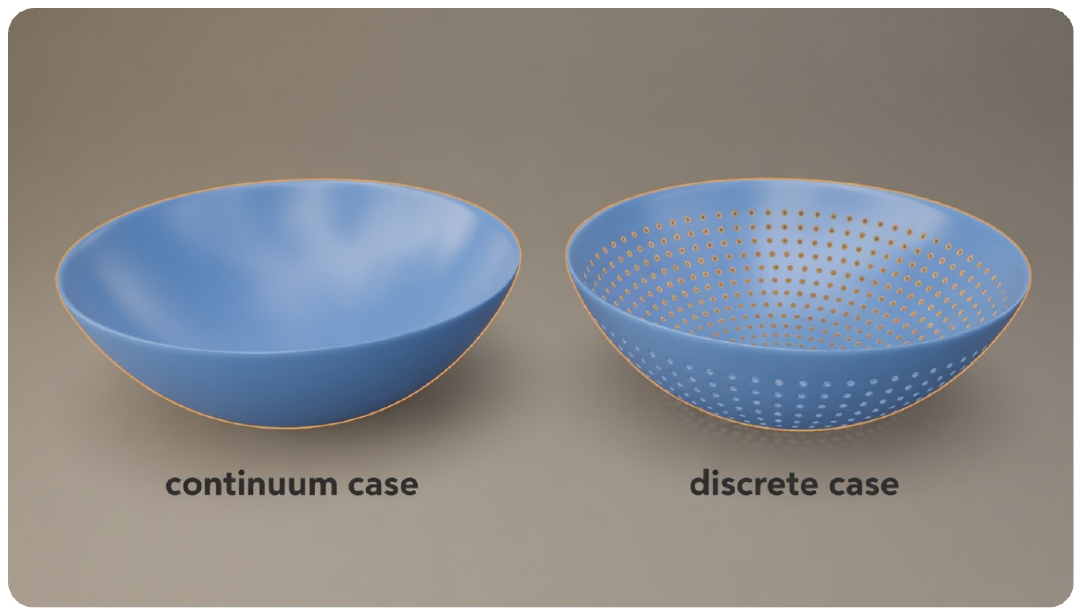}}
\caption{We can imaging the manifold of system parameters in the continuous regime for the collective supermode as a kind of "bowl" that protects quantum correlations from "leakage"—that is, from decoherence processes.
}
\end{figure}

\subsection{The Multiverse Picture and Spectral Geometry of Correlation Invariants}

The geometric framework proposed in this work naturally admits a
hierarchical interpretation.

A particular physical realization of an open quantum system is called a
\emph{Planet}. A Planet is characterized by its Hilbert space,
Hamiltonian, Liouvillian, family of correlation invariants, and the
associated Correlation Landscape.

Changing the physical parameters of the system—including interaction
strengths, graph topology, dissipation rates, external driving,
temperature, or the choice of collective modes—continuously deforms the
Planet and its Correlation Landscape. The collection of all such
physically admissible models forms the \emph{Multiverse}. In this
picture, each point of the Multiverse corresponds to one Planet, while
the evolution of the theory consists not only in the dynamics of quantum
states on a single Planet but also in understanding how Correlation
Landscapes evolve throughout the Multiverse under continuous parameter
deformations.


The central hypothesis of the proposed theory is that Correlation
Blockade is determined by the spectral geometry of the Liouvillian
rather than by individual correlation functions.

Suppose that a chosen family of correlation invariants defines a
submanifold of the density-matrix space through the conditions

\[
\langle K_\alpha\rangle=0.
\]

If this manifold is invariant under the Liouvillian evolution, then the
corresponding correlation constraints remain preserved throughout the
dynamics.

The stability of this invariant manifold is governed by the spectrum of
the Liouvillian. Eigenmodes with vanishing or small real parts describe
stationary or slowly relaxing directions of evolution, whereas
eigenmodes possessing large negative real parts decay rapidly and
suppress deviations transverse to the invariant manifold. Consequently,
the existence and robustness of Correlation Blockade are expected to be
controlled not by individual observables but by the spectral structure
of the Liouvillian itself.


A particularly important quantity is the Liouvillian spectral gap,

\[
\Delta
=
\min_{\lambda_i\neq0}
|\mathrm{Re}\,\lambda_i|,
\]

where $\lambda_i$ denote the eigenvalues of the Liouvillian.

The spectral gap determines the separation between slow invariant
dynamics and rapidly decaying modes.

This motivates the following conjecture.

\paragraph{Gap Protection Conjecture.}

Let $\Delta$ denote the Liouvillian spectral gap.

Then

\begin{enumerate}

\item if $\Delta>0$, there exists a stable invariant manifold supporting
Correlation Invariants;

\item increasing $\Delta$ enhances the robustness of Correlation
Blockade against local and collective noise;

\item as $\Delta\rightarrow0$, the separation between slow and fast
Liouvillian modes disappears, the invariant manifold loses stability,
and the topology of the Correlation Landscape undergoes a qualitative
transition, during which Continents may merge, split, or disappear.

\end{enumerate}

Within this picture, the Liouvillian gap acquires a direct geometric
interpretation. Rather than merely determining relaxation times, it
controls the stability of the Correlation Landscape itself. The gap
therefore acts as a geometric protection parameter governing the
existence and persistence of Correlation Blockade under coherent
dynamics, dissipation, and environmental noise.

Planets possessing Liouvillians with equivalent low-energy spectra exhibit topologically equivalent Correlation Landscapes, even if their microscopic Hamiltonians are completely different.

\section{From Correlation Blockade to Correlation-Protected Manifolds}

The theory of open quantum systems traditionally seeks
\emph{Decoherence-Free Subspaces} (DFS), namely linear subspaces of the
Hilbert space that remain invariant under the action of the Liouvillian
or, equivalently, under the system-environment interaction operators.
When the initial state belongs to a DFS, the subsequent dynamics
preserves the subspace, thereby protecting the encoded quantum
information against specific decoherence mechanisms.

The present work proposes a complementary viewpoint. Rather than
starting from invariant linear subspaces, we begin with a prescribed
family of correlation invariants

\[
\mathcal K=\{K_\alpha\},
\]

and define the corresponding Correlation Blockade through the conditions

\[
\langle K_\alpha\rangle=0,
\qquad
\forall K_\alpha\in\mathcal K.
\]

The set of all pure states satisfying these constraints forms the
\emph{Correlation Blockade Set}

\[
\mathcal M_{\rm CB}
=
\left\{
|\Psi>_B:
\langle K_\alpha\rangle_\psi=0,
\;
\forall K_\alpha
\right\}.
\]

Instead of considering only pure states, we construct the convex hull

\[
\mathcal D_{\rm CB}
=
{\rm conv}
(\mathcal M_{\rm CB}),
\]

which consists of all density matrices obtained as convex combinations
of blockade states (this is the \textit{ansatz method}) and satisfies the Liouville equation $d\rho/dt=L[\rho]$. The set $\mathcal D_{\rm CB}$ defines a
\emph{Correlation-Protected Manifold} in the space of density operators. Here $|\Psi>_B<$ is  blockade state.

The central objective is then shifted from finding protected wave
functions to identifying Liouvillian generators for which the manifold
$\mathcal D_{\rm CB}$ is invariant or asymptotically invariant under the
open-system dynamics. In this formulation, the protected object is no
longer a linear subspace of the Hilbert space but a generally nonlinear
geometric manifold characterized by a common family of vanishing
correlation invariants.

\[
\rho=\sum{|\Psi>_B<\Psi|_B},
\]

This viewpoint naturally generalizes the concept of Decoherence-Free
Subspaces. Indeed, every DFS generates a Correlation-Protected
Manifold, whereas the converse is not generally expected to hold.
Consequently, the DFS framework appears as a particular case of a
broader geometric theory based on invariant correlation manifolds.

That is, we substitute the density matrix into the Liouville equation and check whether it satisfies the stationarity condition. If it does, this density matrix will also satisfy the blockade condition, since—by virtue of the matrix's construction (specifically, its linearity)—each of its elements individually satisfies the blockade condition.

The proposed formulation also provides a natural bridge between
correlation theory, Liouvillian spectral geometry, and quantum
typicality. The invariant manifold is expected to be determined by the
low-lying spectral structure of the Liouvillian, while its robustness is
governed by the Liouvillian spectral gap. In the continuum limit, the
set of blockade states is conjectured to form extended connected regions
(Continents) within the Correlation Landscape. Finite-dimensional
discretization transforms the Bowl into a Colander, introducing gaps in
the accessible manifold due to the finite resolution of the collective
mode representation. Finally, in the high-dimensional limit, the
concentration of measure is expected to suppress collective
fluctuations, rendering the Correlation-Protected Manifold stable
against sufficiently weak local and collective noise.

These observations suggest a new research program in which the primary
objects are no longer individual protected states but invariant
geometric structures in the space of density operators. The principal
problem becomes the classification of Correlation-Protected Manifolds,
their dependence on the Liouvillian spectrum, their topology within the
Correlation Landscape, and their evolution across the Multiverse of
physical models.

=====================================
\section*{Acknowledgments}

The author expresses gratitude to M. T. for valuable assistance on a metaphysical level.

\section{Appendix}
\subsection{Mathematical formalism}

Here, we introduce a simple procedure for calculating the probability amplitudes of stationary states for many-particle systems in the Bose-Hubbard model.

We can write the state of the system $| \psi \rangle$ as the sum of $n-$particle states $| \psi^{(n)} \rangle$, which in turn is a superposition of states $| i j k \rangle$, representing different distributions of $n$ particles over three modes of the system as

\begin{equation} \label{Eqt1_}
| \psi \rangle = \sum_{n\leq M} | \psi^{(n)} \rangle=\sum_{n\leq M}\sum_{i+j+k=n} c_{ijk}| i j k \rangle,
\end{equation}

where $c_{ijk}$ is the probability amplitude for the state $| i j k \rangle$ with $i+j+k=n$, which denotes a state with $i$ particles in the first mode, $j$ particles in the second mode and $k$ particles in the third mode. $M$ is the size of the truncated Hilbert space (the maximum number of particles in the system).

From the standard Bose-Hubbard Hamiltonian, one can derive the following system of differential equations for the state amplitudes $c_{ijk}$.
\begin{equation}\label{Eqt2_}
\begin{split}
i \frac{d c_{ijk}}{dt} = {}& 
\bigl[\Delta_1 i + \Delta_2 j + \Delta_3 k \\
&+ U \bigl(i(i-1) + j(j-1) + k(k-1)\bigr)\bigr] c_{ijk} \\
&+ F_1 \sqrt{i+1}\, c_{(i+1)jk} + F_1 \sqrt{i}\, c_{(i-1)jk} \\
&+ F_2 \sqrt{j+1}\, c_{i(j+1)k} + F_2 \sqrt{j}\, c_{i(j-1)k} \\
&+ F_3 \sqrt{k+1}\, c_{ij(k+1)} + F_3 \sqrt{k}\, c_{ij(k-1)} \\
&+ g_{12} \sqrt{i(j+1)}\, c_{(i-1)(j+1)k} 
 + g_{21} \sqrt{j(i+1)}\, c_{(i+1)(j-1)k} \\
&+ g_{13} \sqrt{i(k+1)}\, c_{(i-1)j(k+1)} 
 + g_{31} \sqrt{k(i+1)}\, c_{(i+1)j(k-1)} \\
&+ g_{23} \sqrt{j(k+1)}\, c_{i(j-1)(k+1)} 
 + g_{32} \sqrt{k(j+1)}\, c_{i(j+1)(k-1)}.
\end{split}
\end{equation}

Note that $n$ identical particles can be placed in $m$ cavities in ${n+m-1 \choose n} = \frac{(n+m-1)!}{n!(m-1)!}$ ways. With $m=3$, one has $T_{n+1}={(n+1)(n+2)}/{2}$ for different combinations, where $T_{n}$ stands for the so-called triangular number.

Following \cite{Wang2021}, we can write the equation governing for the time dependency of the $n$-particle state $| \psi^{(n)}(t) \rangle$:
\begin{equation} \label{Eqt4_}
i\frac{d}{dt}| \psi^{(n)}(t) \rangle \approx H_{0} | \psi^{(n)}(t) \rangle + H_{+} | \psi^{(n-1)}(t) \rangle,
\end{equation}
where we introduced the particle-conserving part of the total non-Hermitian Hamiltonian $H_{0}=\hat{a}^{\dag{}}\mathbf{G_{eff}}\hat{a}+\sum_i{U_{i}\hat{a}^{\dag{}}_{i}\hat{a}^{\dag{}}_{i}\hat{a}_{i}\hat{a}_{i}}$ and the part that increases the number of particles by one: $H_{+}=\sum_i{F_{i}\hat{a}_{i}^{\dag{}}}$. In addition, we neglected the annihilation part of the pump Hamiltonian in the weak pump limit due to sub-leading ordering of the probability amplitudes, $c^{(n+1)} \ll c^{(n)}$ \cite{Wang2021}.

\begin{figure}
\includegraphics[width=0.49\textwidth]{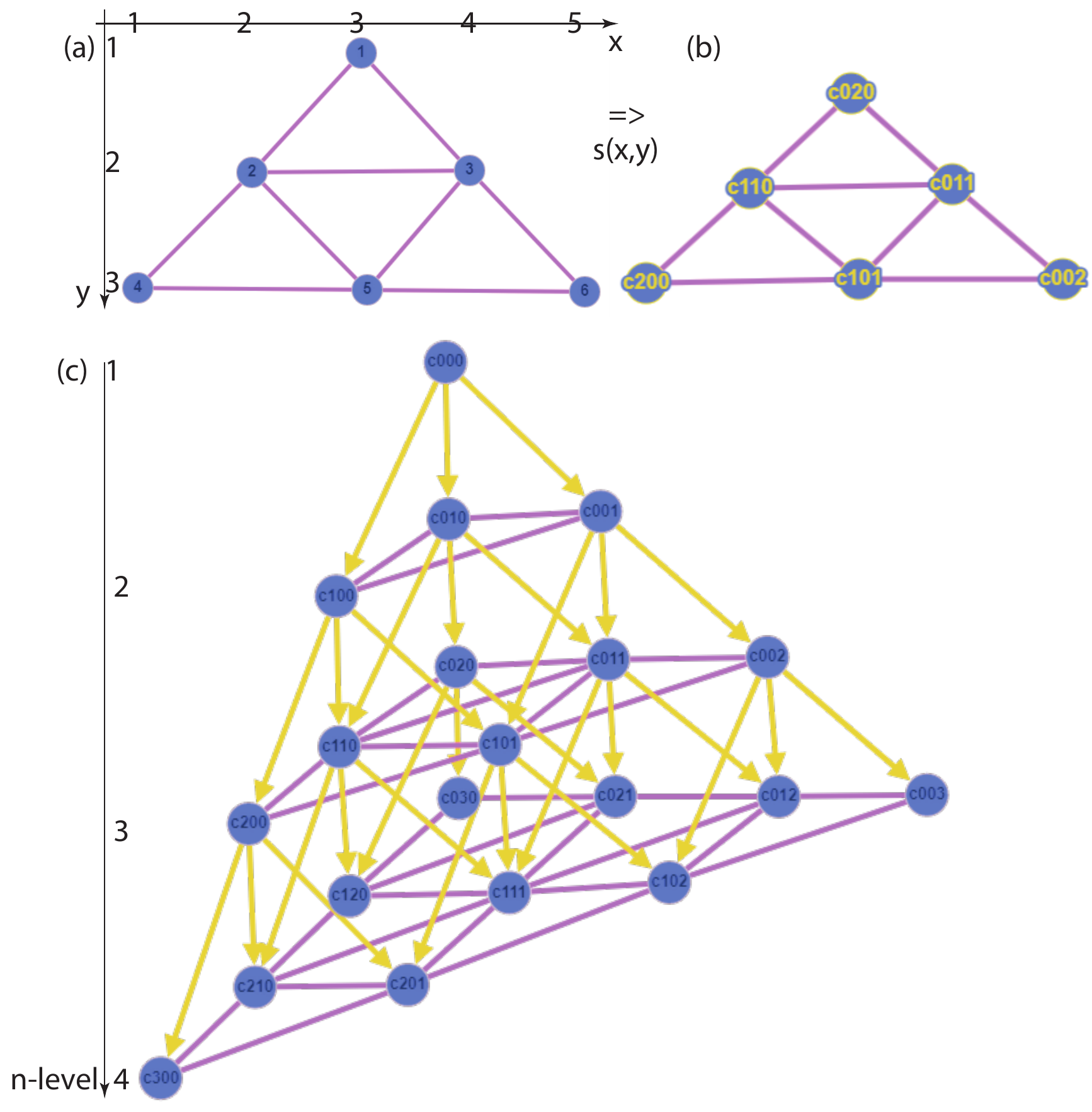}
\caption{(a) The $n$-level triangular graph with $T_{(n+1)}$ with vertices numbered from left to right, top to bottom, each y coordinate corresponds to a certain level at each level $n$ indices are determined by a set of numbers $v^{(n)}$; (b) A sketch of the Fock two-particle tensor state space of a trion.(c) Part of the Fock space of states of the trion. Nodes correspond to different states, graphs with purple edges correspond to $n$-particle tensor states, yellow arrows show transitions between different tensor states.}
\label{figb2}
\end{figure}

From \eqref{Eqt4_}, we can obtain the equation for the $n$-particle state amplitude vector $c^{(n)}$, whose elements $c_{ijk}$ are ordered according to the triangular lattice graph (see Fig.~\ref{figb2}a, from top to bottom, left to right):
\begin{equation} \label{Eqt5_}
i\frac{dc^{(n)}}{dt}=G^{(n)}c^{(n)}+f^{(n)}c^{(n-1)},
\end{equation}
The amplitude vector $c^{(n)}$ contains $T_{n+1}$ elements.

Here we introduce the $n$-particle quadratic form matrix $G^{(n)}$ with size $T_{n+1}$ by $T_{n+1}$. It consists of two matrices:
\begin{equation} \label{EqB6_}
G^{(n)}= \mathcal{D}^{(n)}+\mathcal{G}^{(n)},
\end{equation}
where $\mathcal{D}^{(n)}$ is the diagonal matrix with the following matrix elements:
\begin{equation} \label{EqB7_}
\mathcal{D}^{(n)}_{i,i'}=\sum_{j=1,2,3}{(\Delta_j s(x,y)_j+U s^{(n)}(x,y)_j(s^{(n)}(x,y)_j-1))}\delta_{i,i'},
\end{equation}
here the indices $j=1,2,3$ denote the index of subsystem.
The function $s^{(n)}(x,y)=(y - x, n - y + 1, x - 1)$ takes as input a triangular lattice coordinate function, where the arguments have the following meaning: $x$ is the row number from top to the bottom order starting from 1, $x=\lceil -1 + \sqrt{1 + 8i} \rceil / 2$ and $y=i-\frac{( x - 1 ) \cdot x}{2} $ is the column number counted from left to the right starting from 1. The output of the $s$-function is a triple of numbers $(n_1, n_2, n_3)$ corresponding to the state $|n_1 n_2 n_3 \rangle$ on a given node numbered $i$, see fig.~\ref{figb2}a.

$\mathcal{G}^{(n)}$ in \eqref{EqB6_} is a block matrix that rearranges particles in the $n$-particle state:

\begin{equation} \label{EqB9_}
\mathcal{G}^{(n)}= \begin{bmatrix} \mathcal{G}^{{(n-1)}^{'}} & \mathcal{W}^{(n)} \\ \mathcal{W'}^{{(n)}} & \mathcal{S}^{(n)}\end{bmatrix}.
\end{equation}

An $n-$state graph is obtained from an $(n-1)-$state graph by adding a chain of $n-$states to the bottom side of the triangular graph. The $(n-1)-$state subgraph is described by the matrix $\mathcal{G}^{{(n-1)}^{'}}$, which is also determined recursively through equation \eqref{EqB9_}, but with a modification: $s^{{(n)}'}(x,y) = (y - x, (n+1) - y + 1, x - 1)$.

The $n-$states chain matrix $\mathcal{S}^{(n)}$ consists of the following matrix elements:

\begin{equation} \label{EqB10_}
\mathcal{S}^{(n)}_{ii'}=\begin{cases}
g_{31}\sqrt{s_3^{(n)}(i) s_1^{(n)}(i')},v_i^{(n)}-v_{i'}^{(n)}=1,\\
g_{13}\sqrt{s_1^{(n)}(i) s_3^{(n)}}(i'),v_i^{(n)}-v_{i'}^{(n)}=-1,\\
0,\text{otherwise},
\end{cases}
\end{equation}
here $s_j^{(n)}(i)=s(x(v^{(n)}(i)),y(v^{(n)}(i)))_j$ and $v_i^{(n)}$ is $i$th number site on the $n-$ state chain. This matrix is responsible for the exchange between particles in the first and third subsystems.

Particle exchange matrix between the $(n-1)$ graph and the $n$-chain is defined as the following block matrix:
\begin{equation} \label{EqB11_}
\mathcal{W}^{(n)}=\begin{bmatrix}
  \hat{0}^{(n-2,n)} \\
  R^{(n)}
\end{bmatrix}.
\end{equation}

The sub-matrix $\mathcal{R}^{(n)}$ has the following form:

\begin{equation} \label{EqB11_R}
\mathcal{R}^{(n)}_{ii'}=\begin{cases}
g_{12}\sqrt{s_1^{(n-1)}(i) s_2^{(n)}(i')},v_i^{(n)}-v_{i'}^{(n)}=n-1,\\
g_{13}\sqrt{s_1^{(n-1)}(i) s_3^{(n)}(i')},v_i^{(n)}-v_{i'}^{(n)}=n,\\
0,\text{otherwise}.
\end{cases}
\end{equation}

\begin{equation} \label{EqB12_}
\mathcal{W'}^{(n)}=\begin{bmatrix}
  \hat{0}^{(n-2,n)} \\
  R^{{(n)}'}
\end{bmatrix},
\end{equation}
with the following matrix elements:

\begin{equation} \label{EqB13_}
\mathcal{R'}^{(n)}_{ii'}=\begin{cases}
g_{21}\sqrt{s_2^{(n-1)}(i) s_1^{(n)}(i')},v_i^{(n)}-v_{i'}^{(n)}=n-1,\\
g_{31}\sqrt{s_3^{(n-1)}(i) s_1^{(n)}(i')},v_i^{(n)}-v_{i'}^{(n)}=n,\\
0,\text{otherwise}.
\end{cases}
\end{equation}

In \eqref{Eqt5_} we also introduced a matrix $f^{(n)}$ which transfers the system from the $(n-1)$-particle tensor state to the $n$-particle tensor state of size $T_{n+1}$ by $T_{n}$ with the following matrix elements:

\begin{equation} \label{EqB16_}
f^{(n)}_{ii'}=\begin{cases}
F_2\sqrt{s_2^{(n)}(i)},v_i^{(n)}-v_{i'}^{(n-1)}=0,\\
F_1\sqrt{s_1^{(n)}(i)},v_i^{(n)}-v_{i'}^{(n-1)}=T_{n-1}-1,\\
F_3\sqrt{s_3^{(n)}(i)},v_i^{(n)}-v_{i'}^{(n-1)}=T_{n-1}+1,\\
0,\text{otherwise}.
\end{cases}
\end{equation}

For example, for the two-particle state this matrix looks as follows:
\begin{equation} \label{Eqt9_}
f^{(2)}= \begin{bmatrix} \sqrt{2} F_2 & 0 & 0 \\ F_1 & F_2 & 0 \\ F_3 & 0 & F_2 \\ 0 & \sqrt{2} F_1 & 0 \\ 0 & F_3 & F_1 \\ 0 & 0 &\sqrt{2} F_3 , 
\end{bmatrix}
\end{equation}
while the particle-conserving matrix reads:
\begin{widetext}
\begin{equation} \label{Eqt8_}
G^{(2)}= \begin{bmatrix} 2U + 2 \Delta_2 & \sqrt{2} g_{21} & \sqrt{2} g_{23} & 0 & 0 & 0 \\ \sqrt{2} g_{12} & \Delta_1 + \Delta_2 & g_{13} & \sqrt{2} g_{21} & g_{23} & 0 \\ \sqrt{2} g_{32} & g_{31} & \Delta_2 + \Delta_3 & 0 & g_{21} & \sqrt{2}g_{23} \\ 0 & \sqrt{2} g_{12} & 0 & 2U + 2 \Delta_1 & \sqrt{2} g_{13} & 0 \\ 0 & g_{32} & g_{12} & \sqrt{2} g_{31} & \Delta_1 + \Delta_3 & \sqrt{2}g_{13} \\ 0 & 0 & \sqrt{2} g_{32} & 0 & \sqrt{2} g_{31} & 2U + 2 \Delta_3 \end{bmatrix}
\end{equation}
\end{widetext}

%


Using the resulting matrices $G,f$ we can find the following stationary solutions of the equations \eqref{Eqt5_} to the probability amplitudes $c^{(n)}$ by recurrently solving them for each $n=1,2,3...$:

\begin{equation} \label{Eqt10_}
\begin{array}{l}{c^{(1)}=-G^{{(1)}^{-1}}f^{(1)},}\\
{c^{(2)}=-G^{{(2)}^{-1}}f^{(2)}c^{(1)}=G^{{(2)}^{-1}} f^{(2)} G^{{(1)}^{-1}}f^{(1)},}\\
{...,}\\
{c^{(n)}=- G^{{(n)}^{-1}}f^{(n)}c^{(n-1)}=(-1)^n\prod\limits_{i=n}^{1} G^{{(i)}^{-1}}f^{(i)}.}
\end{array}
\end{equation}


\begin{thebibliography}{99}
\bibitem{1} I. Carusotto and C. Ciuti, Quantum fluids of light, Rev. Mod. Phys. \textbf{85}, 299 (2013).

\bibitem{2} A. Imamoglu, H. Schmidt, G. Woods, and M. Deutsch, Strongly interacting photons in a nonlinear cavity, Phys. Rev. Lett. \textbf{79}, 1467 (1997).

\bibitem{3} H. Flayac and V. Savona, Input-output theory of the unconventional photon blockade, Phys. Rev. A \textbf{88}, 033836 (2013).

\bibitem{4} A. Lingenfelter, D. Roberts, and A. A. Clerk, Unconditional Fock state generation using arbitrarily weak photonic nonlinearities, Phys. Rev. A \textbf{102}, 033707 (2020).

\bibitem{5} E. Zubizarreta Casalengua, J. C. L\'{o}pez Carre\~{n}o, F. P. Laussy, and E. del Valle, Conventional and Unconventional Photon Blockade, Laser Photon. Rev. \textbf{14}, 1900279 (2020).

\bibitem{6} T. A. Khudaiberganov, I. Yu. Chestnov, and S. M. Arakelian, Total Blockade Effect in a Polariton Trimer, JETP Lett. \textbf{122}, 405 (2025).
\bibitem{Wang2021} Y. Wang, W. Verstraelen, B. Zhang, T. C. H. Liew, and Y. D. Chong, \href{https://journals.aps.org/prl/abstract/10.1103/PhysRevLett.127.240402}{Phys. Rev. Lett. 127, 240402 (2021).}
\end{thebibliography}
\end{document}